# Stability Criteria, Atomization and Non-thermal Processes in Liquids

S.D.Kaim

Technology University of Opole, K.Sosnkowskiego 31, Opole, 45-272, Poland

National University of Odessa, Dvorjanskaja 2, Odessa, UA-65026, Ukraine

e-mail: s0x80@mail.ru tel:+38-0482-634898

## Abstract

Analyzing the first equation in the BBGKY chain of equations for an equilibrium liquid-gas system, we derived the analytical expression for the atom work function from liquid into gas. The coupling between the atom work function from liquid into vacuum and the stability criterion of liquid in limiting points of the first type was shown (using I.Z. Fisher classification). As it turned out, Fisher's criterion corresponds to the condition of atomization. We have expressed the state equation in terms of the atom work function from liquid into vacuum and performed calculations of the limiting line of stability composed of limiting points of the first type for argon. Our model discovers an interesting effect of the negative atom work function: at a constant volume of liquid, on a temperature rise (and also at a fixed temperature and decreasing specific volume of liquid) the atom work function drops and takes a negative value with a modulus that is significantly larger than the atomic thermal energy. We propose a new two-stage mechanism of sonoluminescence based on non-thermal processes in liquid in a state with a negative atom work function. The first stage includes the emission of atoms from the interior of the bubble into gas at hyper-thermal velocities. At the second stage a collision of emitted flow takes place between the gas atoms along with the implosion of the central part of the bubble. As a result of the impact excitation, ionization and the subsequent recombination, the flash of electromagnetic radiation that can be seen in sonoluminescence experiments develops.

**Key words:** atom work function, equation of state, stability criteria, fluid atomization, mechanism of sonoluminescence

## 1. INTRODUCTION

Thermo-activation processes play an important role in the forming of the equilibrium and non-equilibrium properties of homogeneous and inhomogeneous condensed systems. These processes define at a microscopic level the realization possibility of phase equilibrium, the direction and evolution behavior of the systems under non-equilibrium conditions. Activation processes are sensitive to the inter-particle interactions in the system, whose characteristics define the behavior of the system near the limiting points and lines of stability on phase diagrams. At the microscopic level, activation processes have mono, duo and multi-particle natures and play a special role when the realization of metastable and absolutely unstable system states takes place. The energetics of these processes is defined by the nature of inter-particle interactions and the effective self-consistent field produced by all the particles of the system. Therefore, in order to investigate the behavior of mono-particle activation processes in condensed systems, the calculations of the self-consistent mono-particle potentials under certain thermodynamic conditions are usually used.

For condensed systems in solid, liquid or gaseous states, we typically find relative stability at varying thermodynamic conditions. Accordingly, metastable and absolutely unstable states can also be found, and one of the significant tasks in statistical physics is to localize and find the nature of the instability points and limiting line of stability in any phase state. The most general microscopic approach for the investigation of limits of stability in liquids, solids and gases is based on the analysis of the Bogolubov-Born-Green-Kirkwood-Yvon (BBGKY) chain of equations [1-3]. For homogeneous liquids and gases the analysis of instability points and limiting lines of stability refers to the investigation of the asymptotic behavior of the radial distribution function and to the exploration of the hard spheres unary distribution function [1-3]. Two-phase equilibrium conditions and coexistence lines are also available to study the analyzing of the first equation of the BBGKY chain of equations [1-3].

One of the mono-particle functions sensitive to the macroscopic parameters of the system is a work function of atoms from system into vacuum, or from one phase into another. The aim of the present work is to investigate the limiting lines of stability for simple liquids and gases using the first equation of the BBGKY chain of equations and the calculations of the atom work function from liquid or gas into vacuum. The analytical expression for the atom work function is derived from the analysis of asymptotes for the potential acting on a separate atom in an inhomogeneous system. The limiting line of stability of the simple liquids is plotted according to the relationship between the stability criterion in the limiting points of the first type and the atom work function from liquid into vacuum established in the present work.

As a possible application of the developed approach and the model calculations of the atom work function, we built up a new double-stage mechanism of sonoluminescence in liquids. The main engine for the sonoluminescence is an emission of atoms with hyper-thermal velocities from the interior of the bubble cavity. Converging on the center of the bubble, the flow of atoms leads to the impact excitation and ionization of the gas atoms with the implosion of the gas in the bubble.

## 2. THE ATOM WORK FUNCTION FROM LIQUID

We take as given the existence of a flat interface in a two-phase simple liquid-gas equilibrium system. The interaction of atoms will be described by means of pair-interaction potential independent of system density. The Hamiltonian of this system may be expressed as

$$H = \sum_{i=1}^{N} \frac{\mathbf{P}_i^2}{2M} + \frac{1}{2}\sum_{i \neq j=1}^{N} \Phi(\mathbf{R}_i - \mathbf{R}_j) \tag{1}$$

where $N$ – the number of atoms in the liquid; $\mathbf{P}_i, M$ – the impulse and mass of the atoms; and $\Phi(\mathbf{R})$– the interaction energy of the pair of atoms. The unary atomic distribution function in inhomogeneous liquid $F_1(z)$ satisfies the first equation of the BBGKY chain of equations [1]

$$k_B T \frac{\partial}{\partial z} F_1(z) + \frac{1}{v_0}\int d^3 R_1 F_2(z, z_1, \rho^{||}) \frac{\partial}{\partial z} \Phi(z, z_1, \rho^{||}) = 0, \tag{2}$$

where $k_B, T$ – the Boltzmann constant and temperature; $v_0$ – the volume per atom in the liquid; $F_2(z, z_1, \rho^{||})$ – the pair distribution function of the atoms; and $\rho^{||} = \mathbf{R}^{||} - \mathbf{R}_1^{||}$ – the constituent of the vectors difference parallel to the surface with equation $z = 0$. The next representation of the pair distribution function $F_2(z, z_1, \rho^{||})$ was used.

$$F_2(z, z_1, \rho^{||}) = F_1(z) F_1(z_1) g(z, z_1, \rho^{||}) \tag{3}$$

where $g(z, z_1, \rho^{||})$ – the pair correlation function of inhomogeneous liquid.

Taking (3) into account, the equation (2) may be written in the equivalent integral form

$$F_1(z) = \exp\left[-\frac{U(z)}{k_B T}\right], \tag{4}$$

where the potential plays the role of the self-consistent potential in the Boltzman distribution produced by the atoms of the liquid in the current point, and which effectively acts on a separate atom

$$U(z) = -\frac{1}{v_0}\int_{-\infty}^{z} dz_1 \int d^3 R_2 F_1(z_2) g(z_1, z_2, \rho_{12}^{||}) \frac{\partial \Phi(R_{12})}{\partial R_{12}} \frac{z_2 - z_1}{R_{12}}. \tag{5}$$

As long as we define the atom work function as the work that needs to be done to move a separate atom from a point in the liquid (the point should be far enough from the interface) into vacuum and leave it there with a velocity equal to zero. It is thus natural to connect the difference in the asymptotic values of the potential (5) with the atom work function.

To evaluate the asymptotic values of self-consistent mono-atomic potential far from the interface, approximations of the pair correlation and unary distribution functions were made.

$$g\left(z, z_1, \rho^{\|}\right) \cong g_0\left(R_{12}\right), \; F_1(z) \cong \Theta(-z) \tag{6}$$

where $g_0\left(R_{12}\right)$,– the pair correlation function of homogeneous liquid; and $\Theta(z)$– the Heviside staircase function. Asymptotic values of mono-atomic potential inside the liquid and in vacuum are correspondingly

$$\lim_{z \to -\infty} U(z) = 0 \tag{7}$$

$$\lim_{z \to \infty} U(z) = \frac{4\pi}{3v_0} \int_0^\infty dR \frac{\partial \Phi}{\partial R} g_0(R) R^3 \tag{8}$$

The difference between the asymptotic values of potential in vacuum (8) and inside the liquid (7) is equal to the atom work function from liquid into vacuum [4]

$$A_{l-v} = \frac{4\pi}{3v_0} \int_0^\infty dR \frac{\partial \Phi}{\partial R} g_0(R) R^3. \tag{9}$$

By analogy, it is possible to derive the expression for the atom work function from gas into vacuum. The difference between the atom work functions from liquid into vacuum and from gas into vacuum is equal to the atom work function from liquid into gas [4]

$$A_{l-g} = \frac{4\pi}{3} \int_0^\infty dR \frac{\partial \Phi}{\partial R} \left[ \frac{1}{v_0} g_0(R) - \frac{1}{v_1} g_1(R) \right] R^3, \tag{10}$$

where $v_1, g_1(R)$– the volume per atom in gas and atom pair correlation function of homogeneous gas.

For the model calculations of the atom work function from liquid into vacuum $A_{l-v}$, from gas into vacuum $A_{g-v}$ and from liquid into gas $A_{l-g}$, we used the Lennard-Jones model potential, with parameters $\varepsilon$ and $\sigma$ taken from [5]

$$\Phi(R) = 4\varepsilon\left[(\sigma/R)^{12} - (\sigma/R)^6\right], \tag{11}$$

and various approximations for the pair correlation function. The pair correlation function for atoms of liquid and gas were calculated using the results of the thermodynamic perturbation theory in the Barker-Henderson (BH) approach [6,7] and Weeks-Chandler-Anderson (WCA) approach [8,9]. The diameter of the hard sphere in the BH approach was taken as

$$d = \int_0^{\alpha} \left(1 - \exp\left(-\Phi(R)/k_B T\right)\right) \qquad (12)$$

where $\alpha$ – the first zero of the atom interaction potential.

The calculations of $A_{l-v}$, $A_{g-v}$ and $A_{l-g}$ were performed for the temperature interval from melting point up to the critical temperature, with the liquid and gas density data taken from [10]. Figures 1 and 2 show the calculation results of the atom work functions for argon with the pair distribution function in BH and WCA approximation. As can be seen, the atom work function is highly sensitive to the choice of approximation model for the pair distribution function in liquid and gas. In the BH approach the atom work functions $A_{l-v}$, $A_{l-g}$ for argon take negative values along almost the whole length of the investigated temperature interval that does not correspond with stability of the liquid-gas interface. The main advantage of WCA approximation over BH approximation lies in a more accurate description of the pair distribution function with respect to the minimum of inter-atom interaction potential. As a result, the position of the first maximum of the pair correlation function and the "softness" of the repulsive part of the inter-atom potential is taken much more precisely into account. In WCA approximation, which takes the inter-atom pair correlations more precisely into account, the atom work functions $A_{l-v}$, $A_{g-v}$, $A_{l-g}$ have positive values (see Fig. 2).

Figure 3 shows the calculation results for the temperature dependence of the atom work function $A_{l-g}$ for argon in WCA approximation with different parameters for the Lennard-Jones potential [5]. For comparison, Figure 3 shows the experimental data of the temperature dependence for evaporation heat [10], which was re-calculated per single atom and expressed in Kelvin. Calculation of the evaporation heat per single atom corresponds to the postulated mono-atomic evaporation mechanism.

The high sensitivity of calculation results for the work function $A_{l-g}$ from the pair potential parameters and from of the pair distribution function, theoretically allows the attainment of better agreement with the evaporation heat data due to the choice of potential parameters and in a more general sense, due to the choice of its form (the Lennard-Jones potential is a model).

## 3. ATOM WORK FUNCTION FROM LIQUID AND THE LINE OF THE LIMITING POINTS OF FIRST TYPE

Using the general analysis of the Bogolubov equation for the radial distribution function of homogeneous liquid in a superposition approximation, while analyzing the asymptotic of the radial distribution function, the analytical criteria of stability of the liquid was derived and the stability of the solutions of the Bogolubov equation were investigated [1,2]. All the possible limiting points of stability for the uniform phase were classified corresponding to their isolation or non-isolation. In accordance with [1,2], the analytical criterion of stability for the limiting points of the first type is expressed as

$$\frac{4\pi}{3v_0 k_B T}\int_0^\infty \Phi'(r) g_0(r) r^3 dr = 1, \qquad (13)$$

where $\Phi'(r)$ – the derivative of potential energy for the central interaction of the atom pair; $g_0(r)$ – the radial distribution function of the homogeneous liquid; and $v_0$ – the volume per atom in liquid.

Comparing expressions (9) and (13), it is easy to see that the analytical stability criterion of the homogeneous liquid for the first type of limiting points can be expressed in the following form

$$A_{l-v} = k_B T \qquad (14)$$

to give it a transparent physical meaning. Implementation of criterion (14) with a given temperature results in equalizing the atom work function from liquid into vacuum with the atom thermal energy. If the temperature is given, then the equality (14) may hold due to the choice of values for the volume $v_0$. The equation (14) is not linear with respect to the $v_0$, because the integrand function strongly depends on $v_0$. To solve equation (14), we used Lennard-Jones potential with parameters $\varepsilon = 124\,\mathrm{K}$ and $\sigma = 3.418\,\text{Å}$, taken from [5], which correspond to argon. The radial distribution function was modeled on the framework of the thermodynamic perturbation theory in WCA approximation.

Figure 4 shows the solution of equation (14) in the $(V,T)$ plane with respect to the specific volume of the fluid $V$. In the same figure the experimental data for the liquid-gas [10] and the liquid-crystal equilibrium lines are also plotted with the experimental data taken from [1].

As can be seen from Figure 4, equation (14) may be satisfied for temperatures less than $215.8\,\mathrm{K}$, where it has two solutions for specific volume. All the points of the line obtained are limiting points of the first type in Fisher's terminology [1,2]. These points are not isolated and produce a curve in the plane $(V,T)$. Above this curve the condition $A_{l-v} < k_B T$ holds and the liquid or gas is absolutely unstable. From a physical perspective the validity of this inequality means that the internal forces of the system are unable to retain separate atoms, which, if appropriate conditions are created, will leave a system due to their thermal motion energy. Thus the stability criterion for the liquid in the limiting points of the first type has a transparent energetic meaning. In the domain of the triple point the line of the limiting points of the first type is situated a little to the left of the experimental equilibrium liquid-crystal curve, but even with a slight increase in temperature it lies completely in the liquid regions. The temperature corresponding to the maximum of the limiting line of stability is higher than the experimental critical temperature of argon. The limiting line of stability intersects the experimental liquid-gas equilibrium line. For the liquid-gas equilibrium line points, which correspond to the gas phase and are situated slightly to the right from the intersection point, $A_{g-v} < k_B T$ holds, which corresponds with the absolute instability of the gas phase.

## 4. THE ENERGETIC MEANING OF THE GENERAL EQUATION OF STATE FOR LIQUIDS AND GASES

Following the correlation function method, the general equation of state for a homogeneous liquid or gas whose atoms interact by means of pair forces can be expressed as [1,2,5,6]

$$\frac{p}{nk_B T} = 1 - \frac{2\pi n}{3k_B T}\int_0^\infty drr^3 g_0(r)\frac{d\Phi(r)}{dr} \tag{15}$$

where $n$ is the density of the number of atoms. Taking the expression for the atom work function from liquid into vacuum into account, equation (15) may be expressed as

$$p = n\left(k_B T - A_{l-v}/2\right), \tag{16}$$

where the non-ideal part of the state equation (the term that holds the inter-atomic interactions) is expressed in terms of the atom work function from liquid into vacuum. This allows us to give an energetic interpretation of the contribution of this term to the pressure and all the quantities connected with the pressure. It should also be noted that both terms in formula (16) are of a mono-particle nature. The first has a kinetic nature and corresponds to the kinetic energy of ideal gas particles. The second term, though having a mono-particle nature, is however conditioned by pair interactions in the system, and its origin is concerned with forming in the liquid of the self-consistent potential and its influence on the system pressure. Of significance here is that this contribution to pressure in the system is directly connected with the difference in the effective mono-atomic potentials inside the system and outside in vacuum.

The marked relationship of the equation of state for a homogeneous liquid or gas with such one-particle characteristics as an atom work function in vacuum, emphasize a deep connection between the one-particle energy level in condensed matter in relation to vacuum and the general stability conditions.

The virial theorem [11,12] takes a special place in the ranks of the most general results of interacting particle mechanics. This theorem defines the conditions that hold the system in a bounded region of space, i.e. in the region of finite motion for particles [11,12]. Under finite motion conditions the average value of the kinetic energy of the system $\overline{T}$ is connected to the average value of the potential energy $\overline{U}$ [11]

$$2\overline{T} = \overline{\sum_a \mathbf{r}_a \frac{\partial U}{\partial \mathbf{r}_a}}, \tag{17}$$

where $U$ is a potential energy of the system as a function of the coordinates of particles.

Equality $A_{l-v} = k_B T$ defines the limiting points of the I-st type. The condition of retaining the particles in a bounded region only due to the internal interactions and without the influence of any external fields may be derived by analogy to (14). On the other hand, as long as we can consider $A_{l-v}$ as an average self-consistent potential that acts to separate particles of the system, then condition (14) may be treated as a special case of the virial theorem. We can rewrite condition (14) as

$$2 \cdot \frac{3}{2} k_B T = 3A_{l-v} \tag{18}$$

where the left part holds the average kinetic energy, and $A_{l-v} = U(\infty) - U(-\infty)$, and $U(z)$ corresponds to the effective one-atomic potential. It becomes clear that if we postulate the homogeneity of the function $U(z)$, then in compliance with the virial theorem, the self-consistent one-atomic potential $U(z)$ at a large distance from the surface behaves as $U(z) \sim 1/z^3$. This dependence corresponds to the asymptotic of the interaction energy of the atom with the semi-bounded body [13].

## 5. THE SPINODAL EQUATION IN TERMS OF THE ATOM WORK FUNCTION

The spinodal $(\partial p/\partial \rho)_T = 0$, splits the $(V, T)$ phase diagram of a liquid into domains of positive and negative values of the derivative $(\partial p/\partial \rho)_T$, where $p$, $\rho$ – pressure and density [14]. Inside the area bounded by the spinodal the homogeneous phase is absolutely unstable and the domain between the spinodal and binodal (the coexistence curve) corresponds to the metastable states of liquid or gas.

Taking (16) into account, we can write the spinodal equation $(\partial p/\partial \rho)_T = 0$ in terms of the atom work function from liquid into vacuum

$$A_{l-v} + n(\partial A_{l-v}/\partial n)_T = 2k_B T\,. \tag{19}$$

We can also express, in terms of the atom work function, all the thermodynamic quantities that depend on pressure, including the isothermal compressibility of the liquid

$$\beta_T = -\frac{1}{V}\left(\frac{\partial V}{\partial p}\right)_T = 2\left(\frac{2k_B T}{V} - \frac{A_{l-v}}{V} + \frac{\partial A_{l-v}}{\partial V}\right)^{-1} \tag{20}$$

As can be seen from (19) and (20), the isothermal compressibility $\beta_T$ becomes divergent along the spinodal line. The isothermal compressibility directly connects with the root-mean-square fluctuations of the number of atoms, thus, when the system state approaches the spinodal, the role of the fluctuations of the number of particles increases.

## 6. THE VOLUMETRIC DEPENDENCE OF THE ATOM WORK FUNCTION FROM LIQUID INTO VACUUM

The mono-particle nature of the atom work function $A_{l-v}(V,T)$ as a function of volume and temperature requires more detailed investigation. Figure 5 shows calculation results for $A_{l-v}(V,T)$ as a function of volume for the various temperatures. As can be seen, this function reaches a maximum when the volume is greater than the equilibrium value. The atom work functions corresponding to the values of equilibrium specific liquid volumes are shown as a line on Figure 5. Also plotted is a line corresponding to the atom work function from liquid into vacuum along the crystal-liquid equilibrium line (the specific experimental volumes are taken from [1]). If the work function values are positive for the specific volumes corresponding to the liquid-gas equilibrium line, then the work function takes negative values for the specific volumes corresponding to the liquid-crystal equilibrium line within a temperature range slightly higher than the triple point temperature.

An interesting fact is that isothermal increase of the specific volume of the liquid on the liquid-gas equilibrium line initially leads to an increase in the work function value, which decreases after reaching a maximum. This implies that in the domain of metastable states, the liquid should initially show evidence of a "strengthening" effect, and the reverse effect afterwards; while on decreasing liquid volume at a constant temperature, the atom work function into vacuum immediately becomes negative with a rather large absolute value. This effect is probably related to the instability that leads to the liquid-crystal phase transition.

## 7. THE ATOM WORK FUNCTION IN THE HIGH TEMPERATURE DOMAIN

The spinodal and the line of the first type limiting points within the plane $(V,T)$ confines the domains of stable, metastable and absolutely unstable states of liquid and gas. Above the level of the limiting points of the first type, all system states are absolutely unstable with respect to atomization of the system into separate atoms. In this domain the atom work function from liquid into vacuum may take negative values. To explore the temperature dependence of $A_{l-v}(V,T)$ in the domain of absolute instability, we performed the calculations of $A_{l-v}(V,T)$ as a function of temperature with a given volume. The results of these calculations for argon, with a volume corresponding to the triple point, are in Figure 6.

As can be seen, the modulus of the negative values of the atom work function are several times higher than the corresponding thermal motion energy. While the temperature increases, the ratio of the absolute value of the atom work function and the thermal motion energy decreases. Therefore, the atom work function in the high temperature domain can be considered as a measure of effective temperature for this system with respect to the atomization of the system into single atoms. It would be interesting to implement experimentally and determine the effective gas temperature of on the disintegration of the system into separate atoms. As can be seen from Figure 6, even at room temperature the difference between the effective temperature corresponding to the work function into vacuum and the thermodynamic temperature may amount to up to the few $T$ .

In the high temperature domain (a few thousands degrees) the behavior of temperature dependence for $A_{l-v}$ does not change, but the role of interaction forces in the total balance of terms that contribute to the system pressure  decreases. As long as in state equation (16) the atom work function into vacuum represent the atoms' interactions' contribution to the pressure, the positive values of $A_{l-v}$ correspond to the effective decrease in the pressure in an ideal gas and negative ones correspond to the increase in pressure in the system compared to the pressure of the ideal gas. In the high temperature domain the contribution of the kinetic term to the gas pressure may become dominant.

In the high-temperature domain under isochoric conditions (with a volume that corresponds to the triple point) the atom work function from argon at a temperature $T = 1000\,\mathrm{K}$ takes value $A_{l-v} = -6026\,\mathrm{K}$, at $T = 2000\,\mathrm{K}$ $A_{l-v} = -10017\,\mathrm{K}$ and at $T = 3000\,\mathrm{K}$ $A_{l-v} = -13069\,\mathrm{K}$.

## 8. A NEW MECHANISM OF SONOLUMINESCENCE IN LIQUIDS

The experimental data and the theoretical works on single bubble sonoluminescence (SBSL) available to date, do not explain the common mechanism and new features of this phenomenon [15]. However, the spectra observed in the experiments implemented in the work [16], definitely show the non-thermal nature of the processes in SBSL.

Using the method for the calculation of the atom work function from liquid developed in the current work, we can qualitatively explain the main properties and the features of the SBSL phenomenon. According to our results, it is probable that the liquid states for the temperature and volumetric dependence of the atom work function from liquid into vacuum $A_{l-v}(V,T)$ can be realized with a negative atom work function. These states correspond to the absolute instability of liquid. The atoms are able to spontaneously leave the liquid with a kinetic energy that is a few times higher than that of the corresponding thermal energy. It should also be noted that these liquid states can even be obtained at room temperature, by means of appropriate choice of the specific volume. The physical reason for the realization of the liquid states with a negative atom work function is an existence in liquid of a self-consistent field that acts on the separate atoms of the system and acts by pair inter-particle interactions and correlations. The level of the self-consistent one-particle potential in the liquid may be both lower and higher than the energy level of a separate atom in vacuum (far enough from the interface). The experimental possibility of obtaining and observing the decay of instable liquid states depends on the ability of obtaining the open surface of a liquid with the volumetric parameters corresponding to such absolutely instable states.

During the generation of cavity bubbles in a periodic acoustic field with the following appearance of a strongly pressed liquid state, the collapse of the bubbles is probably accompanied by the appearance of states with a negative atom work function and an open surface. The large changes in the speed of the bubble wall observed during the collapse of the bubble, when the speed of the wall reaches the speed of sound or exceeds it in a short time, correspond to a large acceleration of the liquid surrounding the bubble. The large accelerations result in high liquid pressure, similar to that found in a shock wave, which under these extreme conditions may arise and reach the liquid-bubble interface. The existence of the open surface at the liquid-gas interface and high pressures in the liquid surrounding the bubble result in the conditions at which negative atom work function states take place.

The occurrence of the above states will necessarily lead to the emission of atoms with hyper-thermal velocities from the internal surface of the bubble. The velocity of the emitted atoms will be significantly greater than the corresponding thermal velocity, the speed of sound, and the speed of the bubble wall. The flow of the emitted atoms will proceed towards the center of the bubble. The subsequent behavior of the flow depends on the density of gas inside the bubble. If the density is low enough, then the spherically symmetrical converging flow of emitted atoms will be able to reach the center of the bubble where cross scattering will take place. In relatively high density gas inside the bubble, gas implosion by a flow of atoms with hyper-thermal velocities will be observed. It is follows that on a head-on collision of two atoms with the equal velocities, the effective energy corresponding to their relative motion will be quadrupled. As a result, the spectrum of the impact-excited atoms will contain temperatures four times higher than the hyper-thermal one. When convergence of spherically symmetrical flows of emitted atoms with the velocities much higher than the speed of the bubble wall takes place, the head-on collision of atoms near the center and a possible formation of bunches of particles will occur before the collapse of the bubble. The effect of forward luminescence in the central part of the bubble on the complete collapse of the bubble has been observed in experiments [15]. It should also be noted that the self-consistent one-atom field in liquid acts as an accelerator for the neutral particles.

As already explained, the surface emission of the converging flow of atoms will occur if states with a negative atom work function appear in the liquid surrounding the bubble. The shock wave that emerges to the surface of the liquid may take the role of an immediate catalyst for the emission of atoms. In this case, the duration of the emission of atoms from the surface of the liquid can easily be evaluated. It is well known that in liquids the width of the shock wave front is defined by an atom free path $l$, with a speed equal to or higher than the speed of sound $v$. The time required for the shock wave to go through the near-surface layer of thickness $l$, where an emission due to the extreme conditions at the front of the shock wave will occur, is $\tau = l/v$. To evaluate this numerically, the following numbers should be assumed: $l = 2 \cdot 10^{-8}$ m, $v = 1000$ m/s. Accordingly $\tau = 2 \cdot 10^{-11}$ s. The duration of the light flashes experimentally observed in sonoluminescence are of the same order. The flow of atoms converging to the center of the bubble with a duration $\tau$ is able to generate the light flash, as a result of impact excitation of the atom electron shells and of ionization by collision with the following recombination and a possible thermalization of the group of particles. The extremely short duration of the light flash in experiments on sonoluminescence can be understood from this. The extremely short duration of the flash had been a serious argument against the thermal nature of the light observed in experiments for a long time [15]. The proposed shock-wave mechanism does not conflict with the short duration of the light flash and with the nonthermal origin of the spectra of $Ar$ and the ionized states $O_2^+$ observed in experiments [16]. The impact mechanism of atom excitation and ionization are also in accordance with experimental data. If the pair collisions prevail at the center of the bubble , then the extremely large impact energy of the atoms becomes possible. Therefore, the extremely high excitation energies of $Ar$ atoms observed in experiments [16] are explaned by a pair of head-on $Ar$ atom collisions in a converging flow of atoms. It should also be noted that as long as squeezing and heating take place in a shock wave at the same time, the modulus of the atom work function from liquid at the front of the shock wave may attain quite high values.

If the duration of the emission of atoms from the bubble surface is $\tau$ and the modulus of the atom work function is $|A_{l-v}|$, then the width $d$ of the converging flow of atoms yields $d = \tau\sqrt{2|A_{l-v}|/m}$, where $m$ is the mass of the emitted atom. For the flow of argon atoms at $\tau = 20\,\text{ps}$ and $|A_{l-v}|/k_B = 3000\,\text{K}$, $d = 2.26 \cdot 10^{-8}\,\text{m}$ is obtained. If pair collisions mainly take place in the converging flow of atoms near the center of the bubble, then the size of the central luminous part must be in the order of $d$. However, the converging flow of atoms may lead either to a single scattering or a multiple scattering of atoms onto each other and of atoms onto ions. In the event of multiple scattering of hyper-thermal atoms, the size of the region where the effective excitation and ionization of atoms will take place will be a few time larger than the width of the convergent flow $d$. It should also be recalled that under high-density gas conditions inside the bubble, the probability of the gas scattering of emitted atoms before they reach the center, becomes greater. That explains why at certain densities the impact mechanism may be insufficient to start the gas implosion in the center of the bubble. Thus sonoluminescence, in a high-density region, should have a threshold value on a density parameter. However, if the concentration of gas is too low, the emission mechanism may also be insufficient for effective sonoluminescence, while in a region with low concentrations of gas, sonoluminescence will either be absent or will also have a threshold character. The experimental observations of sonoluminescence show fairly thin intervals in gas concentration at wich it take place in liquid [15].

One more feature that can be explained within the framework of the impact excitation mechanism of electron degrees of freedom and atom ionization on implosion of gas is the duration of the light flashes, which are the same in the various spectrum bands [15]. The flashes in the various frequency ranges take place simultaneously to the generation of impact-excited electron degrees of freedom, and only over the times of generation. The observed shift of the emitted light spectrum into the ultraviolet region shows that the process of light generation is strongly non-equilibrium. Thus the observed results correspond to a non-equilibrium distribution of energy in spectrum, with the level of imbalance strongly dependent on the emission process at the bubble surface. The high level of nonequilibrium of the implosion processes requires consideration within the framework of the kinetic description.

## 9. CONCLUSIONS

The present work shows the meaning and significance of the atom work function from liquid or gas into vacuum for the better understanding of absolute instability phenomena in liquids or gases. The derived relationships for the absolute instability line (the line of the first type of limiting points) and the spinodal equation, allow us to better understand the nature of the physical phenomena and the behavior of the matter on the boundaries of absolute instability. This work gives the atom work function an implicit physical meaning and expresses the thermodynamic parameters of the system in terms of an energetic characteristic.

The marked relationship between the equation of state for the homogeneous liquid or gas with such mono-particle characteristics as the atom work function into the vacuum, allows us to emphasize the deep connection of mono-particle energy levels in the condensed medium with the vacuum and general stability conditions of this medium. That is why the further investigation of the work function and its relationship with other characteristics is an important constituent of the stability problem.

In the high-temperature region, the calculated temperature dependence of the atom work function from liquid or gas into vacuum indicates significant negative values for the work function. Furthermore, the distinct difference in the absolute value of the atom work function into the vacuum, and the thermal energy of atoms, would be interesting to check experimentally.

Within the frameworks of the double-stage mechanism of sonoluminescence developed in the current work, and taking into account the available experimental data, we analyzed the main features of this phenomenon. An important role in sonoluminescence is played by the converging to the center of the bubble flow of atoms emitted from the interface. The emission of atoms from the open surface of the liquid is possible only when the liquid is in a negative atom work function state. These states appear during the strongly non-equilibrium conditions in the liquid-gas system, for example on strong shock pressure. A shock wave emerging to the liquid-gas interface may become an immediate catalyst for the emission of atoms from the surface of the liquid with hyper-thermal velocities. The next stage includes the implosion of the gas at the center of the cavity bubble with the resultant flow of emitted atoms and the ensuing excitation of electronic degrees of freedom and impact ionization of atoms. The subsequent "highlighting" of the excited degrees of freedom in atoms and the recombination of electrons and ions, will lead to the generation of the radiation impulse.

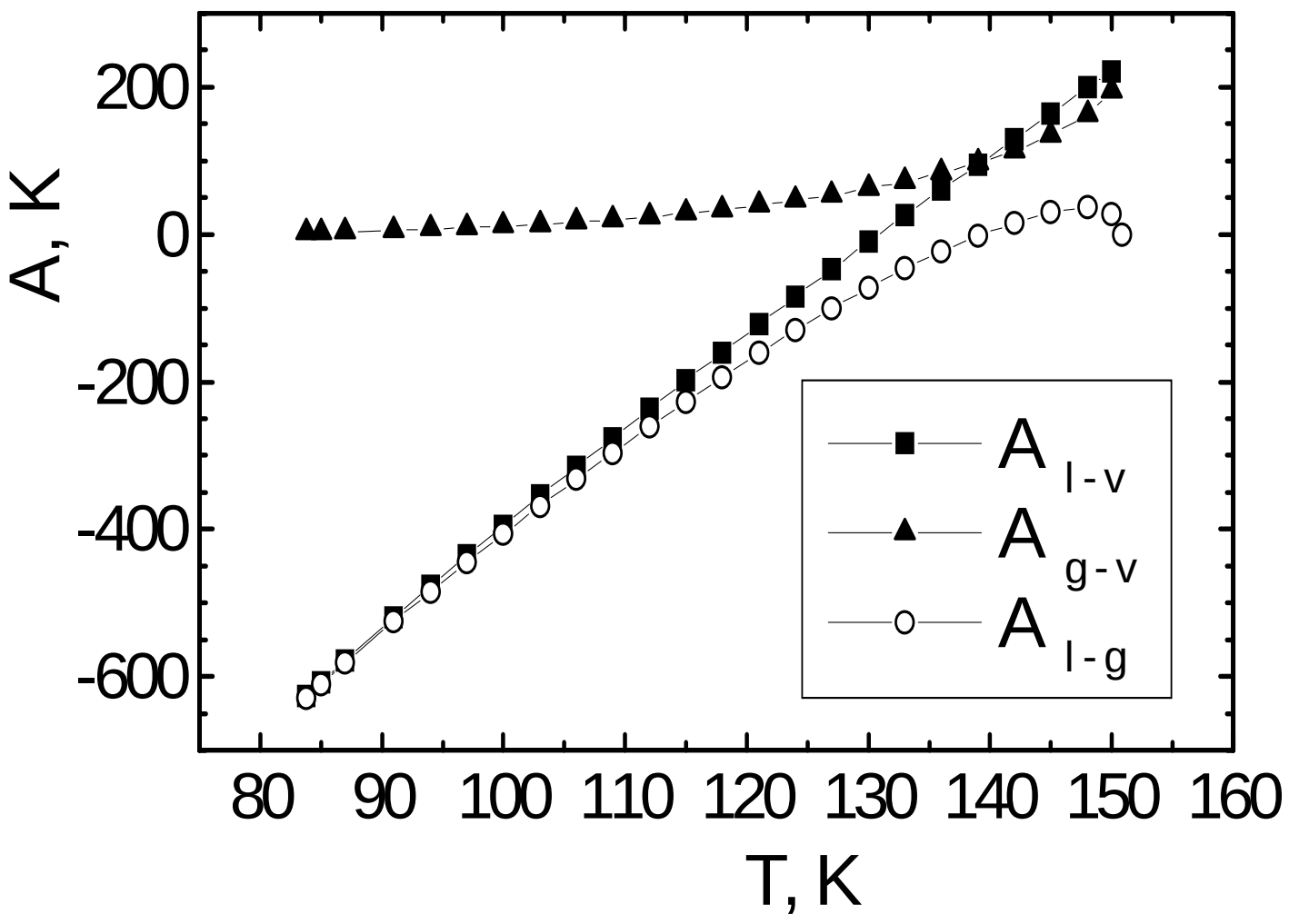

Figure 1. The calculation results of the temperature dependence of the atom work functions $A_{l-v}$, $A_{g-v}$, $A_{l-g}$ for argon in BH approximation. Potential parameters $\varepsilon = 124\,\text{K}$, $\sigma = 3.418\,\text{Å}$.

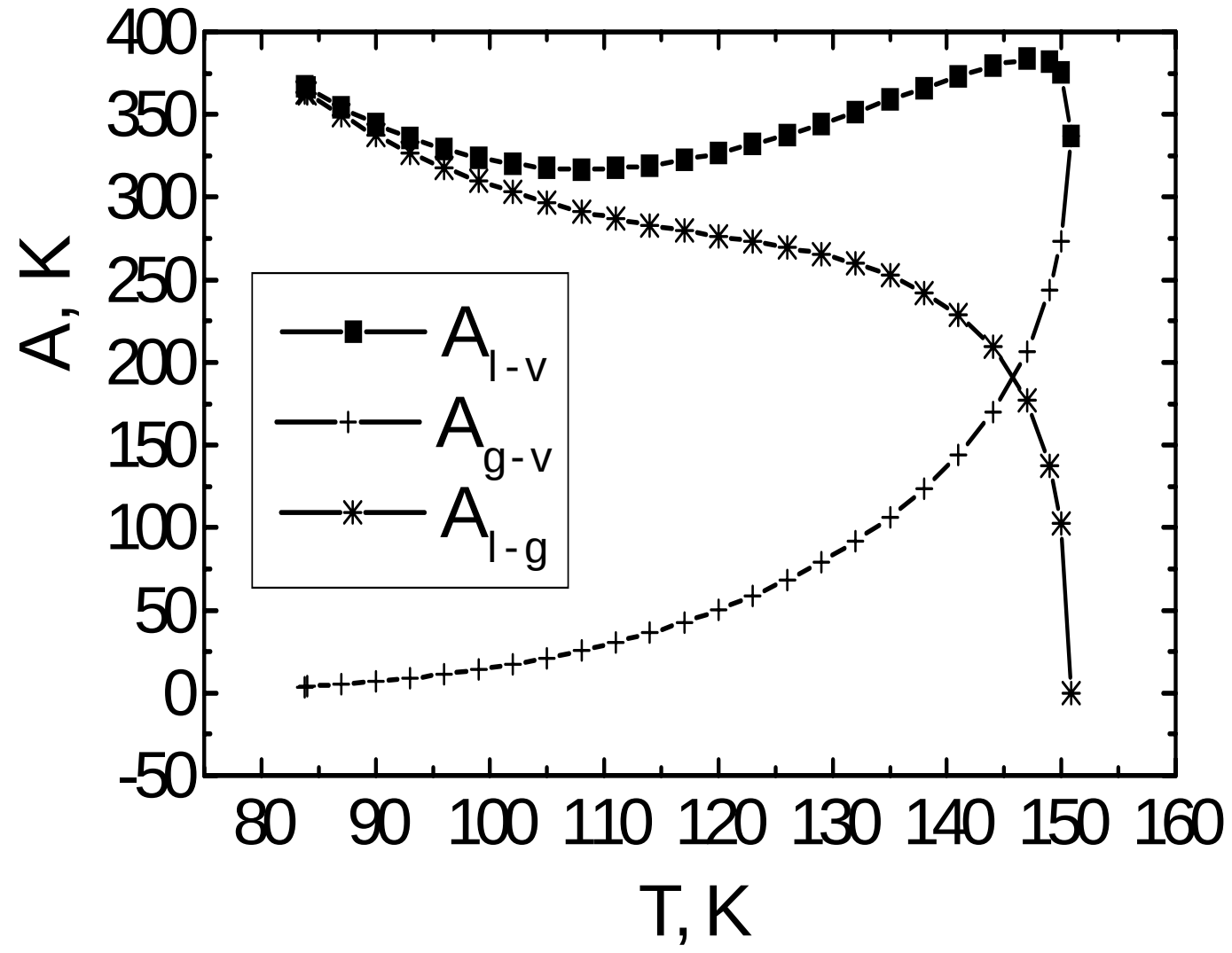

Figure 2. The calculation results of the temperature dependence of the atom work functions $A_{l-v}$, $A_{g-v}$, $A_{l-g}$ for argon in WCA approximation. Potential parameters $\varepsilon = 124\,\text{K}$, $\sigma = 3.418\,\text{Å}$.

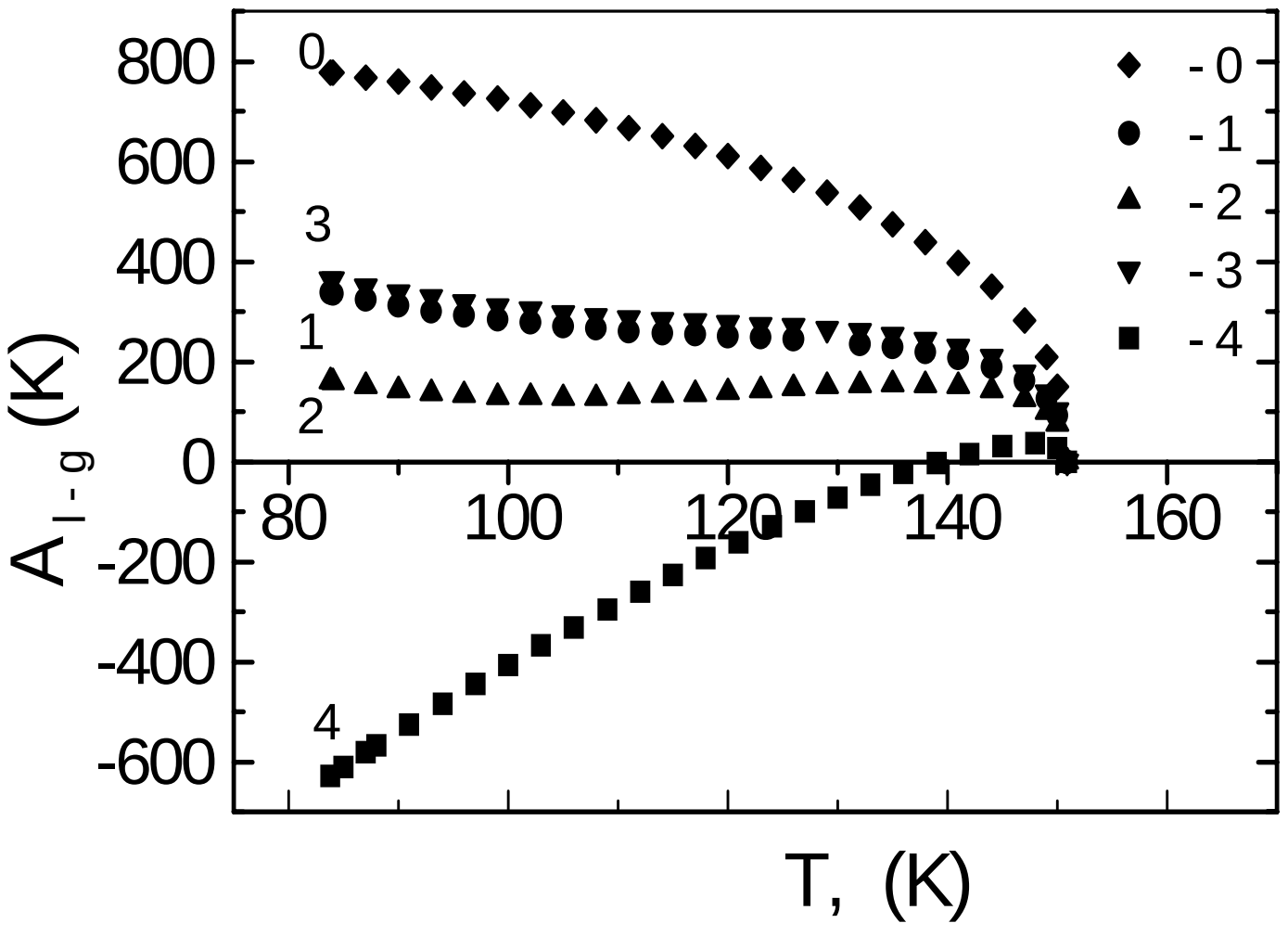

Figure 3. The calculation results of the temperature dependence of the atom work functions $A_{l-g}$ for argon in BH and WCA approximations. Diagram 0 – experimental data; 1 – calculation with $\varepsilon = 119.8\,\mathrm{K}$, $\sigma = 3.405\,\text{Å}$ (WCA); 2- calculation with $\varepsilon = 116\,\mathrm{K}$, $\sigma = 3.465\,\text{Å}$ (WCA); 3 - calculation with $\varepsilon = 124\,\mathrm{K}$, $\sigma = 3.418\,\text{Å}$ (WCA); 4 - calculation with $\varepsilon = 124\,\mathrm{K}$, $\sigma = 3.418\,\text{Å}$ (BH).

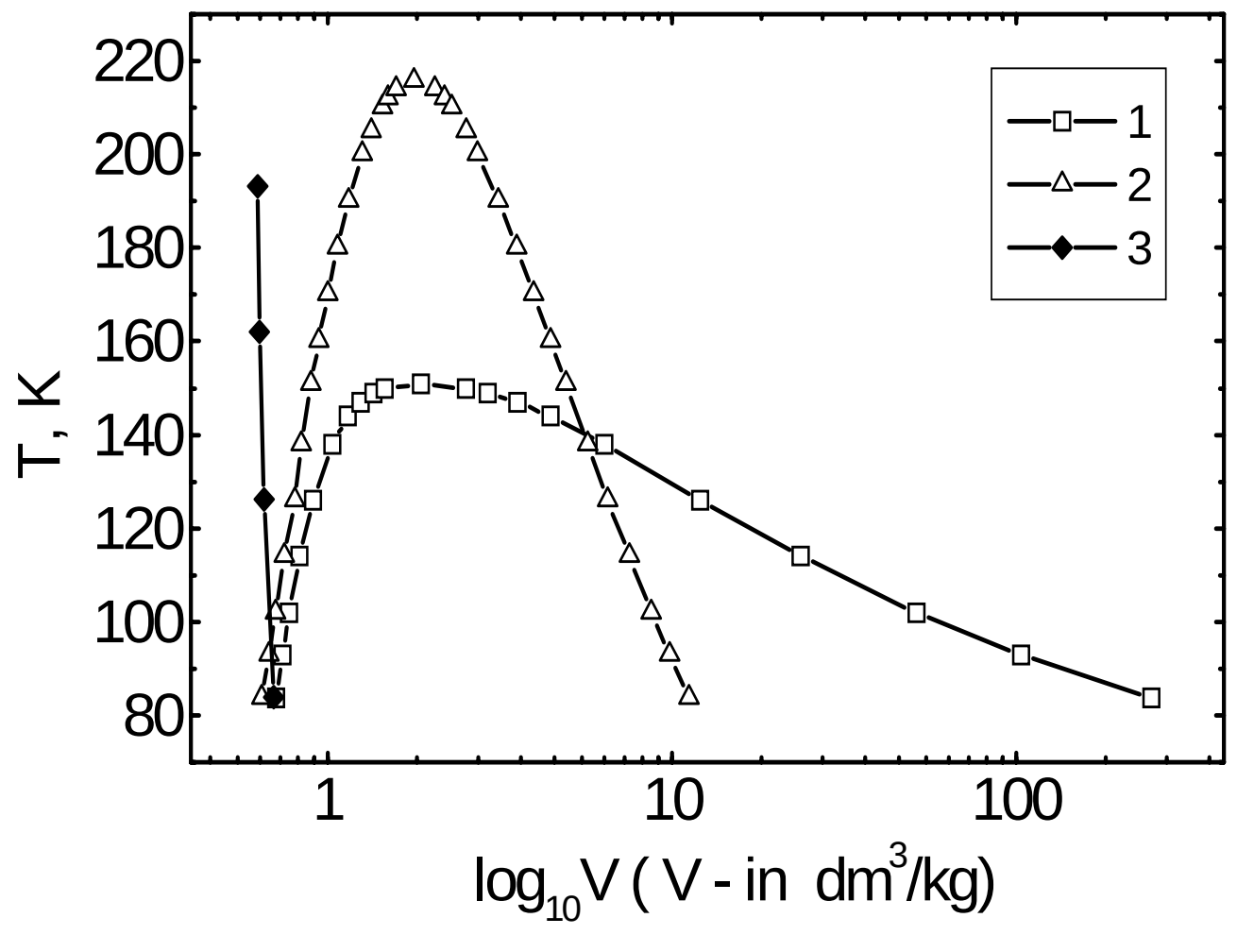

Figure 4. The calculation results of the line of I-st type limiting points for argon. Potential parameters $\varepsilon = 124\,\mathrm{K}$, $\sigma = 3.418\,\text{Å}$. Diagram 1 – equilibrium line of the liquid–gas coexistence (experiment); 2 - first type limiting points line (calculation); 3 - equilibrium line of the liquid-crystal coexistence (experiment).

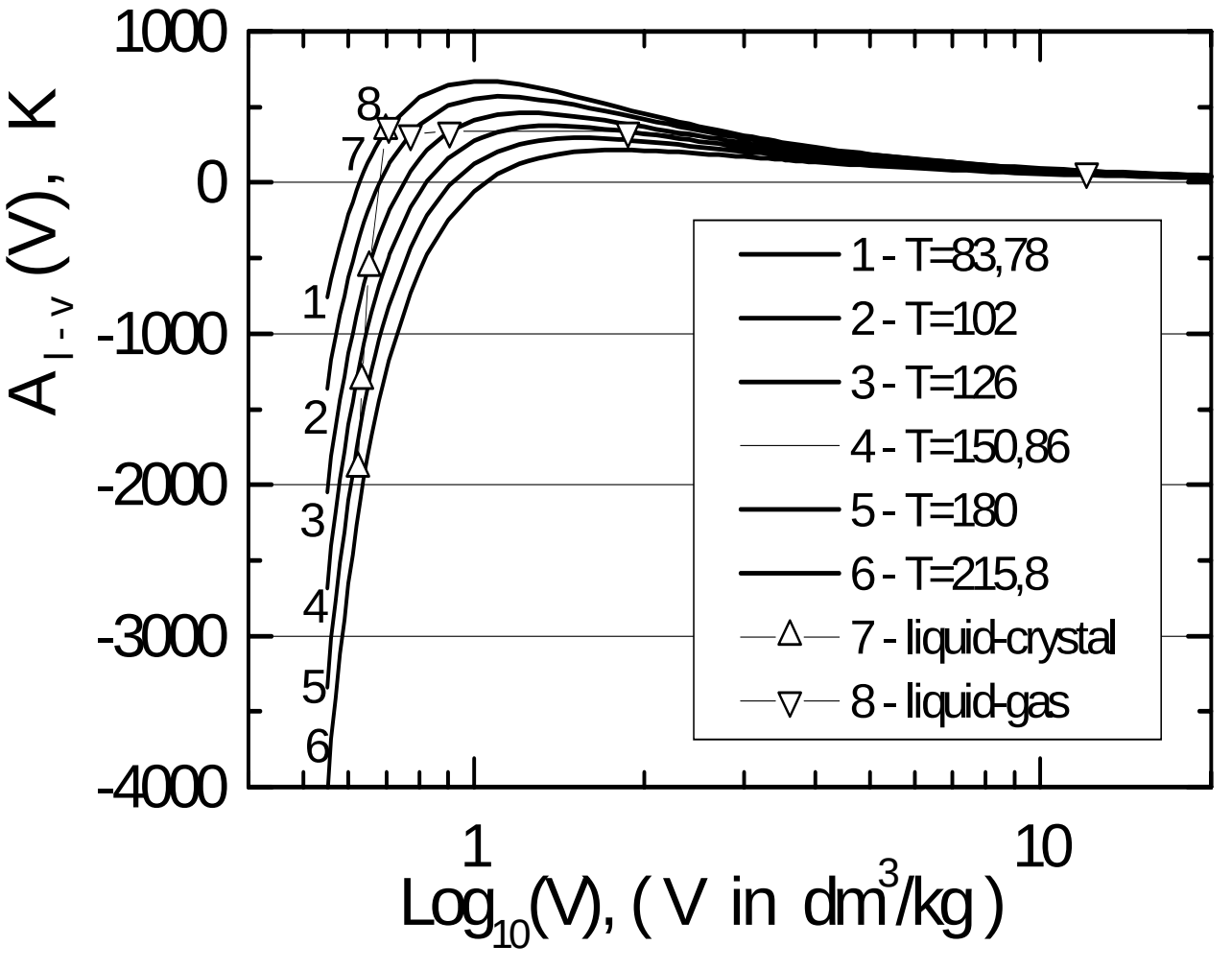

Figure 5. Volumetric dependence of the atom work function $A_{l-v}$ for the different temperatures.

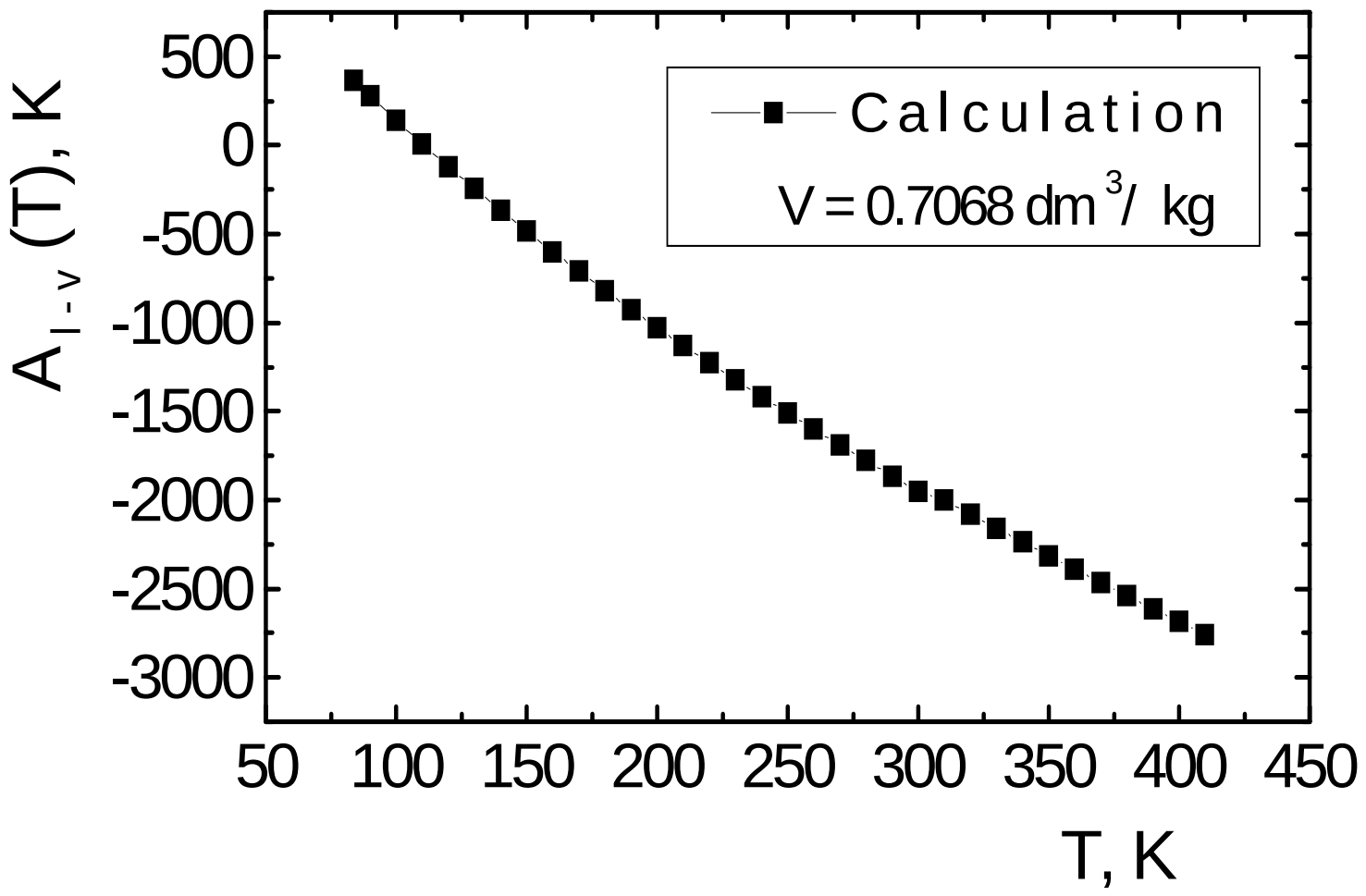

Figure 6. Temperature dependence of the atom work function $A_{l-v}$ for the fixed specific volume of argon ($V = 0.7068\ \mathrm{dm^3/kg}$).

# Literature

1. I.Z.Fisher, Stasistical Theory of Liquids, The University of Chicago Press, Chicago and London, 1964.

2. I.Z.Fisher, Zhurnal Eksperimentalnoi i Teoreticheskoi Fiziki, 28 (1955) 171 (in Russian);
I.Z.Fisher, Zhurnal Eksperimentalnoi i Teoreticheskoi Fiziki, 28 (1955) 437 (in Russian).

3. S.V.Tyablikov, Zhurnal Eksperimentalnoi i Teoreticheskoi Fiziki, 17 (1947) 386 (in Russian).

4. Ya.S. Kaim, Ukrainian Journal of Physics, 49 (2004) 174.

5. J.O.Hirschfelder, C,F.Curtiss, R.B.Bird, Molecular Theory of Gases and Liquids, Wiley, New York, 1954.

6. C.A.Croxton, Liquid State Physics. A Statistical Mechanical Introduction, Cambridge University Press, Cambridge, 1974.

7. J.A.Barker, D.Henderson, Journal of Chemical Physics, 47 (1967) 2856.

8. H.C.Andersen, J.D.Weeks, D.Chandler, Physical Review A, 4 (1971) 1597.

9. L.Verlet, J.J.Weis, Physical Review A, 5 (1972) 939.

10. B.Vargaftik, Handbook on Thermal-Physical Properties of Gases and Liquids, Nauka, Moscow, 1972 (in Russian).

11. L.D.Landau, E.M.Lifshitz, Mechanics, Nauka, Moscow, 1988 (in Russian).

12. V.D.Shafranov, Uspechi Fizicheskich Nauk 128 (1979) 161 (in Russian).

13. E.M.Lifshitz, L.P.Pitaevskii, Statistical Physics, Part 2, Nauka, Moscow, 1978 (in Russian).

14. V.P. Skripov, Metastable Liquid, Nauka, Moscow, 1972 (in Russian).

15. M.P.Brenner, S.Hilgenfeldt, D.Lohse, Reviews of Modern Physics, 74 (2002) 425.

16. D.J.Flannigan, K.S.Suslick, Nature, 434 (2005) 52.